\documentclass[vecphys]{svmult}

\usepackage{makeidx}         
\usepackage{graphicx}        
\usepackage{multicol}        
\usepackage[bottom]{footmisc}

\makeindex             

\begin{document}

\title*{Life on the fast lane: the burst mode at the VLT at present 
and in the future}
\titlerunning{The burst mode at the VLT}
\author{
Andrea Richichi\inst{1} \and
Octavi Fors\inst{2} \and
Elena Mason\inst{3} \and
Marco Delb{\'o}\inst{4} \and
J{\"o}rg Stegmaier\inst{1} \and
Gert Finger\inst{1}
}
\authorrunning{A. Richichi et al.}
\institute{European Southern Observatory, Garching, Germany
\texttt{arichich@eso.org}
\and Departament d'Astronomia i Meteorologia, 
Barcelona, Spain
\texttt{ofors@am.ub.es}
\and European Southern Observatory, Paranal, Chile
\texttt{emason@eso.org}
\and Observatoire de la C{\^o}te d'Azur, Nice, France
\texttt{delbo@oca.eu}
\and European Southern Observatory, Garching, Germany
\texttt{jstegmai@eso.org}
\and European Southern Observatory, Garching, Germany
\texttt{gfinger@eso.org}
}
%
%
\maketitle

\begin{abstract}
The recent implementation of the high-speed 
burst mode at the ISAAC instrument on UT1,
and its propagation to other ESO instruments, has opened the door to 
observational capabilities which hold the potential for a wealth of
novel results. In the ELT era,  when the accent will likely be on lengthy
programs aimed at the best
sensitivity and angular resolution, the VLT telescopes could continue to
play a significant and largely unique role by performing routinely observations
of transient events at high temporal resolution. 
In our contribution, we provide details on
two such kinds of observations, namely 
lunar occultations of stars and of asteroids.
For the first ones, we report on two passages of the Moon in 
regions with high stellar density as the Galactic
Center. The VLT-UT1 telescope was used for the first time to record
successfully 53 and 71 occultations on
March 22 and August 6, 2006, with an angular resolution of 
0.5-1
milliarcsecond and $K\sim12.5$ limiting magnitude. We note that the angular
resolution is superior to that achieved at present by
Adaptive Optics on any telescope, and also superior to that foreseen for the ELT
at the same wavelength. 
LO are also very efficient in terms of telescope time.
We present some of the results, 
including the discovery of close binaries,
and  the detection and study of
compact circumstellar components 
of cool giants, AGB stars and embedded IR sources.

Concerning asteroidal occultations, we aim at observations starting in P80
which would permit high-accuracy, direct determinations
of asteroid sizes for bodies larger than $\approx$50~km. This is
a critical information to
improve our understanding of the physical properties of these
bodies. It will allow us an independent, crucial calibration of the
indirect techniques commonly used to derive estimates of asteroid
sizes and albedo, namely radiometry
(Harris \& Lagerros~\cite{harris})
and polarimetry (Cellino et al \cite{cellino}, and references therein). 
Lunar occultations can be used also to detect
asteroid binary systems, which have been found recently to be not
very rare. Binary systems are invaluable to estimate asteroid
masses and densities, parameters that are at present very poorly
known. 
\end{abstract}

\section{Scientific motivations}
\label{sec:1}
\begin{figure}
\centering
\includegraphics[width=11cm]{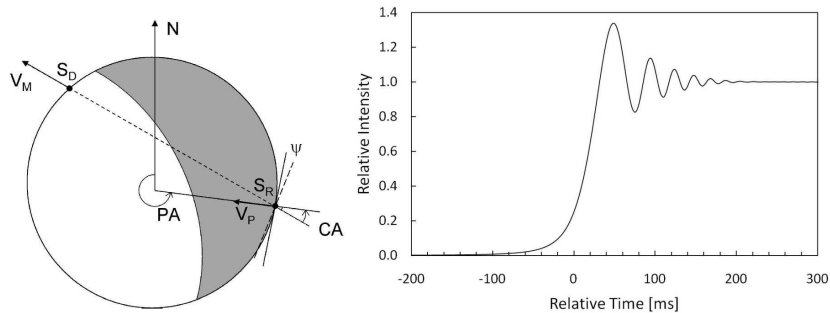}
%
%
\caption{
{\it Left:} Scheme of a lunar occultation event. 
Symbols are explained in
the text.
{\it right:}  A typical (near-IR)
model occultation light curve. Note the temporal
scale of the diffraction fringes, which requires high speed
photometry for proper sampling.
} 
\label{fig_lo}       
\end{figure}

Our main scientific driver are lunar occultations of stellar sources,
and we provide here
some details on the goal and the technical requirements. 
Later we will mention
other astrophysical topics which also require fast, or very
fast, photometry. Lunar occultations (LO) represent a very powerful and yet
extremely simple method to obtain high angular resolution information
on sources covered by the Moon during its apparent motion as seen from
a specific site. The obvious disadvantages of LO are that we cannot
choose the targets at will, and that they are fixed time events.
Fig.~\ref{fig_lo} illustrates the basic geometry of an occultation event.
The position angle (PA) defines the scan direction of the lunar limb
across the source. The local
limb slope $\psi$ can influence this by a few degrees. 
The contact angle CA defines the effective speed
V$_{\rm P}$
of the limb motion, which can be significantly slower than the apparent
lunar motion 
V$_{\rm M}$, especially for LO approaching grazing conditions.
The parameter
V$_{\rm P}$, in conjunction with the observing wavelength, 
defines the rate of the diffraction fringes.
The model light curve of
Fig.~\ref{fig_lo} has been generated for a typical speed of 0.5\,m/s,
in a broad-band K filter.
%

The detailed analysis of the diffraction light curve of a LO is
outside the scope of this contribution, but we mention that it is
rather simple and relatively quick. It
can recover information on the occulted source with angular
resolutions of $\approx$1\,mas, and can even be used to reconstruct
brightness profiles of complex sources without any model-dependent
parameters. These characteristics make LO competitive with
other high-angular resolution methods such as AO or long-baseline
interferometry. However, the light curve must be sampled at
millisecond rates.
This was easily achieved with simple photometers: single pixel
InSb detectors have been routinely used for this work in the near-IR.
On the other hand, these photometers presented the disadvantage
of integrating the light over the whole diaphragm, thus resulting
in added noise from the rather high background normally encountered
during LO observations.

The progressive introduction of array detectors
has significantly changed the landscape of near-IR instrumentation.
Nowadays, almost all telescopes of medium to large size are equipped
with such detectors, which emphasize area and read-out noise (RON) but,
with the possible exception of 
specialized detectors for wavefront sensing, are usually
slow to read-out. However, suitable compromises
can be realized by reading out fast only parts of the arrays.
A comparison of the two approaches was made by
Richichi~\cite{1997IAUS..158...71R}, and we refer to this paper for an
analytic and quantitative comparison of the signal-to-noise ratio (SNR)  for the
photometer and the array cases under various combinations of telescope size,
source brightness and background intensity. The comparison 
included 4\,m and 8\,m telescopes: while
these latter might have seemed a remote possibility ten years ago,  LO have now
been observed on one of the 8.2\,m VLT telescopes equipped with the ISAAC
instrument in the so-called burst mode (Richichi et al.
\cite{2006Msngr.126....24R}). 

Presently, the burst mode is limited to sampling times of 3-5\,ms. This is
suitable for LO work, as well as for other applications of moderately fast
photometry such as occultation and transit phenomena in the solar system,
transient and oscillatory phenomena in stars and binary stars, and some
aspects of the highly time variable properties of 
white dwarfs, neutron stars and pulsars - where of course sensitivity
is the main barrier. A list of high-time resolution astrophysical
topics and available instruments
has recently been presented by Redfern \& Ryan
\cite{redfern}.

\subsection{The burst mode of ISAAC}
\label{sec:2}
A fast readout
 mode on a user-defined subarray has recently been
implemented, tested and commissioned on the Aladdin detector of the
ISAAC instrument \cite{2006Msngr.126....24R}. 
This burst mode is optimized for speed, at the expense of data
organization which needs to be performed offline. The raw data consist
in a sequential write of successive reads
of the subarray, i.e. they are double the size of the standard FITS
cubes and do not have a FITS structure. 
A similar mode, so-called Fastjitter, can achieve standard
data organization but is several times slower.
The main characteristics of the burst mode are listed in 
Table~\ref{tab_burst}.

\begin{table}
\centering
\caption{Main parameters of the ISAAC burst mode}
\label{tab_burst}       
%
%
\begin{tabular}{cccc}
\hline\noalign{\smallskip}
Window & Field of & Min DIT & Max \\
Size (px) & View ($"$) & (ms) & NDIT \\
\noalign{\smallskip}\hline\noalign{\smallskip}
32x32& 4.7x4.7 & 3.2 & 16000\\
64x64& 9.5x9.5 & 6.4 & 16000\\
128x128& 19x19 & 14 & 4000\\
256x256& 38x38 & 37 & 1000\\
\noalign{\smallskip}\hline
\end{tabular}
\end{table}

\begin{figure}
\centering
\includegraphics[width=11cm]{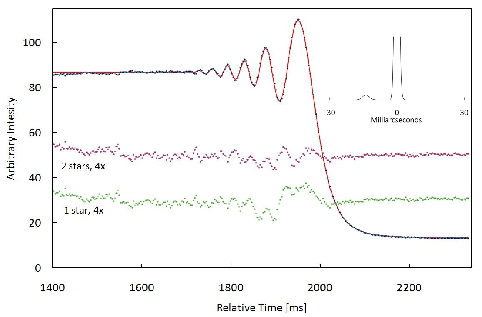}
%
%
\caption{
An anonymous star observed on August 6, 2006
 (2MASS 17524903-2822586),
shown to be a small separation, low contrast binary.
The main panel shows the data, the best fit with a binary
star, and two sets of fit residuals:
one for a single source and one for the binary model.
The residuals are offset by arbitrary amounts and enlarged for
clarity.
The inset is a profile reconstruction using a model
independent method (Richichi \cite{cal}). The companion has a projected
separation of $12.8\pm2$\,mas, and a brightness ratio of 1:35
(magnitudes of the two components are K=5.23 and K=9.12, respectively).
} 
\label{fig_res}       
\end{figure}

We have used the burst mode on the occasion of two series of lunar
occultations in crowded regions near the galactic center, on March 22
and August 6, 2006. In particular for the second event, we were able
to observe 71 occultations over a period of few hours. For experienced
observers, it is possible to offset the telescope, start the observation,
record the data in a total of about two minutes. 
Preliminary results from these
runs have been reported by Fors et al. \cite{fors_sea}. Among them,
we mention 7 small separation binaries, 5 angular diameters, and
4 sources with extended circumstellar emission. An example is shown
in Fig.~\ref{fig_res}.

The study by 
Richichi~\cite{1997IAUS..158...71R} predicted that an 8\,m telescope 
equipped with a near-IR array detector would
reach between K=12 and 14\,mag, depending on the lunar phase and background,
with an integration time of 12\,ms at SNR=10. The preliminary results with ISAAC
reported by Richichi et al.~\cite{2006Msngr.126....24R} show a limiting
magnitude K$\approx$12.5 at SNR=1 and 3ms integration time. Within the
uncertainty of the lunar background, the results are in perfect agreement with
the decade-old prediction. 
Thanks also to the introduction of massive predictions based on
IR survey catalogues such as 2MASS and of automated data pipelines
based on new methods of light curve characterization based on the
wavelet theory (Fors et al. \cite{fors_aa}) it is now possible
to imagine programs of routine LO observations also at the VLT.
We have prepared one such program as a filler for brief unused telescope
times, which has been approved for P80 and has been resubmitted
for P81.

\subsection{Asteroid occultations}
The interest in LO of asteroids stems from the possibility 
to derive the sizes of these bodies from the duration of the occultation 
event. We also note that for objects with angular sizes of 
$\approx$20\,mas or larger, the LO light curves can be analyzed
by simple geometrical (rather than diffraction) optics.
In these cases,
details of the brightness profile, connected with the
distribution of albedo and irregularities on the surface, can be measured. 
Moreover, the technique is ideally suited to discover new binary systems.
Asteroids are moving objects and the prediction of the epoch and geometry 
of their lunar occultations, along with the visibility of the events from
 Paranal, must be calculated for each of object. The K$\approx$12.5 
limiting magnitude
 of the high-speed burst mode of ISAAC on the UT1, roughly scales to a 
limiting asteroid magnitude of V$\approx$14.5. 
Using a computer code to predict the lunar occultations involving the first
 5,000 numbered asteroids (asteroids with number $>5,000$ are in general 
likely to have V$>$14.5), we find that there are about 40 lunar occultations 
of asteroids observable from Paranal every year.  
The unique combination of ISAAC and UT1 makes it possible to observe
occultations events of asteroids with diameters in the range between
100 and 50\,km in the Main Belt and obtain direct determination of their sizes.
In this size range, 
asteroids display angular extensions between 50 and 100\,mas.
At the average lunar proper motion their disappearance last 
100~ms (minimum). By tuning the integration time to yield SNR$>$10 the 
sizes of these bodies can be determined with 
the unprecedented accuracy of some percent.

%



\printindex
\end{document}